\documentclass[10pt]{iopart}

\newcommand{\be}{\begin{equation*}}
\newcommand{\ee}{\end{equation*}}
\newcommand{\bea}{\begin{eqnarray*}}
\newcommand{\eea}{\end{eqnarray*}}
\newcommand{\bl}{\begin{align}}
\newcommand{\el}{\end{align}}

\newcommand{\araa}{ARA\&A}		
\newcommand{\apj}{ApJ}			
\newcommand{\apjl}{ApJLett}		
\newcommand{\aap}{A\&A}			
\newcommand{\mnras}{MNRAS}		
\newcommand{\prd}{Phys.~Rev.~D}		
\newcommand{\ssr}{Space~Sci.~Rev.}	
\newcommand{\nat}{Nature}		

\usepackage{iopams}  
\usepackage{epstopdf}
\usepackage{graphicx}
\usepackage{citesort}

\usepackage{changes}
\usepackage{enumerate}

\begin{document}

\title{Relativistic MHD modeling of magnetized neutron stars, pulsar winds, and their nebulae}

\author{\underline{L.~Del Zanna}$^{1,2,3}$, A.~G.~Pili$^{1,2,3}$, B.~Olmi$^{1,2,3}$, N.~Bucciantini$^{2,1,3}$, E.~Amato$^{2,1}$}
\address{$^1$ Dipartimento di Fisica e Astronomia, Universit\`a di Firenze, Italy}
\address{$^2$ INAF - Osservatorio Astrofisico di Arcetri, Firenze, Italy}
\address{$^3$ INFN - Istituto Nazionale di Fisica Nucleare, Sezione di Firenze, Italy}
\ead{luca.delzanna@unifi.it}

%

\begin{abstract}
Neutron stars are among the most fascinating astrophysical sources, being characterized by strong gravity, densities about the nuclear one or even above, and huge magnetic fields. Their observational signatures can be extremely diverse across the electromagnetic spectrum, ranging from the periodic and low-frequency signals of radio pulsars, up to the abrupt high-energy gamma-ray flares of magnetars, where energies of $\sim 10^{46}\,\mathrm{erg}$ are released in a few seconds. Fast-rotating and highly magnetized neutron stars are expected to launch powerful relativistic winds, whose interaction with the supernova remnants gives rise to the non-thermal emission of pulsar wind nebulae, which are known cosmic accelerators of electrons and positrons up to PeV energies. In the extreme cases of proto-magnetars (magnetic fields of $\sim 10^{15}$~G and millisecond periods), a similar mechanism is likely to provide a viable engine for the still mysterious gamma-ray bursts. The key ingredient in all these spectacular manifestations of neutron stars is the presence of strong magnetic fields in their constituent plasma. Here we will present recent updates of a couple of state-of-the-art numerical investigations by the high-energy astrophysics group in Arcetri: a comprehensive modeling of the steady-state axisymmetric structure of rotating magnetized neutron stars in general relativity, and dynamical 3-D MHD simulations of relativistic pulsar winds and their associated nebulae. 
\end{abstract}

%
%
%
%
\ioptwocol

\section{Introduction}

Neutron Stars (NSs) are extremely compact objects, with masses of $M = 1-2\, M_\odot$ confined within radii of $R = 10 - 15$~km, where densities even above the nuclear one are reached. Gravity is so strong in these objects that their internal structure must be governed by Einstein equations (they are generally referred to as {\it relativistic stars}, regardless of their particular composition or equation of state, EoS). NSs were predicted to be the central remnant of supernova explosions  \cite{Baade:1934} right after the discovery of the neutron, and to power young supernova remnants like the Crab nebula  \cite{Pacini:1967} just before the actual identification of NSs as radio pulsars \cite{Hewish:1968}. The engine for emission is the spindown of the compact object, producing electromagnetic and pair plasma winds from a rotating magnetosphere anchored at the star surface. Since the early days of pulsars more than 2000 sources have been discovered in radio, while more recently hundreds of NSs are discovered by the FERMI satellite at the other extreme of the electromagnetic spectrum, namely in $\gamma$-rays, providing new insight on their magnetospheric structure and on the particle acceleration mechanisms \cite{Cerutti:2016,Harding:2016}.

The emission properties of NSs are thus mainly characterized by the presence of strong magnetic fields, typically $B\sim 10^{12}$~G in the magnetosphere for a standard radio pulsar, produced by field amplification during the core collapse of the progenitor star. Another class of NSs, that of {\it magnetars} \cite{Mereghetti:2015,Turolla:2015}, is expected to host even stronger fields, say of $B\sim 10^{16-18}$~G at birth, as under peculiar conditions the initial ones can be further amplified via dynamo processes or magneto-rotational instabilities during the proto-NS phase \cite{Duncan:1992,Moesta:2015}, probably also in the case of merger events \cite{Giacomazzo:2015,Kiuchi:2015}. Magnetars are mainly observed through the sporadic huge flares occurring in their magnetospheres, most probably due to fast reconnection in current sheets \cite{Lyutikov:2006a,Del-Zanna:2016a}, giving rise to {\it Soft Gamma-ray Repeaters} events. Moreover, millisecond proto-magnetars are believed to power long and short {\it Gamma-Ray Bursts} \cite{Metzger:2011,Bucciantini:2009,Bucciantini:2012}, and probably even peculiar double-peaked events via quark deconfinement \cite{Pili:2016}. Finally, the field evolution in the first phases of magnetars could lead to orthogonalization of the spin and magnetic axes, and to a detectable emission of gravitational waves for stars where the magnetic energy density is so dominant such to deform its overall structure \cite{DallOsso:2009}.

In spite of the importance of an accurate modeling of magnetized and rotating relativistic stars, a comprehensive investigation in general relativity was missing until recently, due to the mathematical complexity of the problem. In section~2 we report the latest achievements in this field by the high-energy astrophysics group in Arcetri \cite{Pili:2017}. 

Another important aspect of pulsar physics is the interaction with the environment. Fast-rotating young NSs are known to efficiently drive relativistic outflows of particles (pairs and maybe ions) and electromagnetic fields, so that when impacting on the slowly expanding supernova ejecta a bubble of hot plasma shining in synchrotron light can be observed. Compared to other supernova remnants, this class of so-called {\it Pulsar Wind Nebulae} (PWNe) is precisely characterized by their non-thermal emission from the central regions, produced by electrons and positrons accelerated to ultra-relativistic velocities at the wind {\it Termination Shock} (TS). Due to its vicinity ($\sim 2$~kpc) and relatively young age ($\sim 1000$ years), the Crab nebula is considered as the PWN prototype, it is certainly the best observed object of its class, and probably the most studied astrophysical source beyond the Solar System. To date, more than one hundred PWNe have been discovered in our Galaxy and they are among the most luminous sources of the sky in the high-energy bands of emission, X and $\gamma$-rays \cite{Gaensler:2006,Hester:2008,Buhler:2014,Kargaltsev:2015}.  

From a theoretical point of view, PWNe represent fantastic astrophysical laboratories, namely for the dynamics of relativistic plasmas, for high-energy conversion mechanisms, and for particle acceleration up to extreme (PeV) energies. They are the best investigated example of {\it cosmic accelerators}, and almost certainly the principal antimatter factories in the Galaxy. Moreover, PWNe serve as close and well observed benchmarks to test models for similar physics encountered in other classes of sources, like the engines of gamma-ray bursts or active galactic nuclei. 

The Arcetri group has a long tradition of research in this field, dating back to the pioneering work of Pacini, up to the recent state-of-the-art 3-D relativistic MHD simulations \cite{Olmi:2016,Del-Zanna:2017}. In section~3 we report and discuss some aspects of our most updated results in this extremely fascinating research field of plasma astrophysics.

\section{Magnetized rotating relativistic stars}

In the present section we describe a systematic method to build numerical equilibria for static and rotating magnetized NSs in general relativity. First we present the general equations governing equilibria in stationary and axisymmetric spacetimes, in both the force-free electrodynamics (FFE) and relativistic magnetohydrodynamics (GRMHD) regimes. In order to simplify the mathematics involved, the metric will be assumed to be conformally flat, which has been proved to be a very reasonable and helpful approximation when treating stationary and axisymmetric equilibria of general relativistic plasmas. Einstein equations greatly simplify to a hierarchical set of partial differential equations (PDEs), of the same kind of those governing the electromagnetic structure. The code for building models of magnetized NSs is freely available at \texttt{http://www.arcetri.astro.it/science/ahead/XNS}, whereas the detailed modeling for different magnetic field and rotation configurations can be found in \cite{Pili:2014,Bucciantini:2015,Pili:2015,Pili:2017}.

\subsection{Stationary and axisymmetric equilibria of general relativistic plasmas}

The problem of finding stationary and symmetric equilibria of relativistic plasmas, and in particular of both static and rotating compact stars and/or of their magnetospheres, has been addressed by many authors, starting from the pioneering work by \cite{Goldreich:1969} in flat spacetime. In general relativity the so-called $3+1$ formalism was first employed for electrodynamics of compact objects by \cite{Thorne:1982}, while see \cite{Gourgoulhon:2011} for the most general covariant approach. Applications of GR to the modeling of compact stars have been addressed, among others, by \cite{Bocquet:1995,Kim:2005,Kiuchi:2008,Ciolfi:2009,Frieben:2012,Ciolfi:2013}. In the following we expound our formalism leading to the equations actually employed for the numerical modeling of NSs.

Assume a stationary and axisymmetric {\it circular spacetime} with $\partial_t=\partial_\varphi=0$ of $3+1$ form
$$
ds^2 = - \alpha^2 dt^2 + \gamma_{11} (dx^1)^2 + \gamma_{22} (dx^2)^2 + R^2(d\varphi - \omega dt)^2,
$$
where $\alpha$ is the {\it lapse function}, $\mathbf{\beta} = - \omega R \mathbf{e}_{\hat{\varphi}}$ is the {\it shift vector} (frame dragging velocity), and we have renamed $\gamma_{33}=R^2$. We have assumed a diagonal spatial metric for simplicity, for which both cylindrical and spherical-like coordinates are allowed (we use the hat symbol to identify orthonormal components). The stationary Maxwell equations are
$$
\mathbf{\nabla} \times (\alpha \mathbf{E} + \mathbf{\beta} \times\mathbf{B}) = 0, \quad \mathbf{\nabla} \cdot \mathbf{B} = 0,
$$
$$
\mathbf{\nabla} \times (\alpha \mathbf{B} - \mathbf{\beta} \times\mathbf{E}) = 
\alpha\mathbf{J} - \rho_e \mathbf{\beta}, \quad \mathbf{\nabla} \cdot \mathbf{E} = \rho_e,
$$
where the $\nabla$ refers to the {\it poloidal} coordinates $x^1,x^2$ alone. From the solenoidal constraint we define the {\it magnetic flux function} $\Psi \equiv A_\varphi$ such that
$$
\mathbf{B} = \frac{\mathbf{\nabla}\Psi}{R} \times \mathbf{e}_{\hat{\varphi}} + 
\frac{\mathcal{I}}{\alpha R}\mathbf{e}_{\hat{\varphi}}, \quad(\mathbf{B}\cdot \mathbf{\nabla} \Psi =0 ),
$$
in which we have written $\mathcal{I}=B_{\hat{\varphi}}/(\alpha R)$.
Any function $f$ satisfying $\mathbf{B}\cdot \mathbf{\nabla} f =0$ is constant on the magnetic surfaces $\Psi=\mathrm{const}$, where fieldlines are located, and $f=f(\Psi)$ alone. The last two Maxwell equations provide the conduction current
$$
\mathbf{J} = \frac{\mathbf{\nabla}\mathcal{I}}{\alpha R} \times \mathbf{e}_{\hat{\varphi}} + 
\frac{R}{\alpha} \left[ -  \nabla\cdot \left( \frac{\alpha}{R^2} \nabla\Psi \right)
+ \mathbf{E} \cdot\nabla\omega \right] \mathbf{e}_{\hat{\varphi}}  ,
$$
whereas the first Maxwell equation implies $E_\varphi=0$ and, using $\Phi \equiv A_t$, the electric field can be written as
$$
\alpha \mathbf{E} + \mathbf{\beta} \times\mathbf{B} = \mathbf{\nabla}\Phi \Rightarrow
\mathbf{E}  = \frac{ \mathbf{\nabla}\Phi + \omega \mathbf{\nabla}\Psi }{\alpha}.
$$
In case vacuum solutions are looked for, $\rho_e = \mathbf{J}=0$ and two coupled PDEs for $\Psi$ and $\Phi$ are derived and must be solved in order to derive the electromagnetic configuration \cite{Pili:2017}. Notice that in the case of rotating pulsar magnetospheres, {\it vacuum gaps} with $\mathbf{E}\cdot\mathbf{B}\neq 0$ appear \cite{Goldreich:1969,Michel:1979}. On the other hand, inside a highly conducting plasma the condition  $\mathbf{E}\cdot\mathbf{B}=0$ must hold, then $\mathbf{B}\cdot \mathbf{\nabla}\Phi =0 \Rightarrow \Phi=\Phi(\Psi)$ and a {\it drift velocity} $\mathbf{v}=v \mathbf{e}_{\hat{\varphi}}$ can be always defined such that, if we let $\Omega (\Psi) \equiv - d\Phi/d\Psi$, we may write
$$
\mathbf{E} = - \mathbf{v} \times \mathbf{B} = - \frac{v}{R}\mathbf{\nabla}\Psi,
\quad v = \frac{\Omega-\omega}{\alpha} R.
$$
When the contribution of matter is negligible, for example when modeling the magnetosphere or the wind of magnetized NSs, the Lorentz force $\mathbf{L}= \rho_e \mathbf{E} + \mathbf{J} \times \mathbf{B}$ acting on the plasma dominates and must balance itself. The FFE condition is thus simply $\mathbf{L}=0$ and leads to an extension of the Grad-Shafranov (GS) equation (e.g. \cite{Del-Zanna:1996a} and references therein):
$$
 \nabla \! \cdot \! \left[ \frac{\alpha}{R^2} \left( 1 \! - \! v^2 \right) \! \nabla\Psi \right]
+ \frac{v}{R} \frac{d\Omega}{d\Psi}| \nabla\Psi |^2 + 
\frac{\mathcal{I}}{\alpha R^2}\frac{d\mathcal{I}}{d\Psi} = 0 ,
$$
where it is found that $\mathcal{I}=\mathcal{I}(\Psi)$. This is a PDE providing the magnetic structure $\Psi$  for given $\mathcal{I}(\Psi)$ and $\Omega(\Psi)$, with extra conditions at the {\it light cylinder} $v=1 \Rightarrow R=R_\mathrm{L} \equiv \alpha/(\Omega-\omega)$. The so-called {\it pulsar equation} is retrieved in flat space ($\alpha=1$, $\omega =0$) and $\Omega=\mathrm{const}$. Only a few semi-analytical works present solutions of the GS equation above  \cite{Contopoulos:1999,Gruzinov:2005,Timokhin:2006}, due to the difficulties of applying continuity conditions at the light cylinder, while time-dependent simulations are able to overcome this difficulty by relaxing to the final steady-state solution \cite{Komissarov:2006,McKinney:2006b,Spitkovsky:2006,Kalapotharakos:2009,Tchekhovskoy:2013,Petri:2016}.

Consider now matter (ideal fluid or plasma), for example in order to model the NS interior. Assuming a purely azimuthal flow velocity and uniform rotation $\Omega\equiv u^\varphi/u^t = \mathrm{const}$ we may write
$$
u^\mu = ( \Gamma/\alpha, 0, 0, \Omega\Gamma/\alpha), \quad \Gamma = (1-v^2)^{-1/2}, 
$$
where $v$ retains the previous form, but being now a physical velocity obviously satisfies $v<1$ so that no critical surfaces are expected in the governing equation. By making the simplifying assumptions:
\begin{itemize}
\item barotropic EoS $p=\mathcal{P}(\rho)$ (e.g. polytropic law): $\quad p=K\rho^{1+1/n}$,
\item conservative Lorentz force with potential $\mathcal{M}$: $\quad \mathbf{L} = \rho h \mathbf{\nabla} \mathcal{M}$,
\end{itemize}
the Euler equation (including the Lorentz force)
$$
\rho h a_\mu + \partial_\mu p + u_\mu u^\nu\partial_\nu p = L_\mu \Rightarrow 
\partial_i p  - \rho h \partial_i\ln (\Gamma/\alpha)  = L_i
$$
provides the simple GRMHD Bernoulli integral
$$
\ln (h/h_c) + \ln (\alpha/\alpha_c) - \ln\Gamma  = \mathcal{M} - \mathcal{M}_c,
$$
in which the $c$ label refers to the star center. Further compatibility conditions that must be imposed are $\mathcal{I}=\mathcal{I}(\Psi)$, $\mathcal{M}=\mathcal{M}(\Psi)$, so the GS equation becomes
$$
\nabla \! \cdot \! \left[ \frac{\alpha}{R^2\Gamma^2}  \nabla\Psi \right]
 + \frac{\mathcal{I}}{\alpha R^2}\frac{d\mathcal{I}}{d\Psi} + \alpha \rho h \frac{d\mathcal{M}}{d\Psi}= 0,
$$
with the previous case being retrieved by letting $\rho\to 0$. As far as the magnetic configuration is concerned, the shape of the poloidal field is selected by choosing the magnetization function $\mathcal{M}(\Psi)$, while the toroidal component is related to the current function $\mathcal{I}(\Psi)$. 

\subsection{Einstein equations in conformally flat metric and the \texttt{XNS} code}
 
The stationary and circular $3+1$ metric of the previous section is conveniently approximated assuming the so-called {\it Conformally Flat Condition} (CFC) \cite{Wilson:2003}:
\be
ds^2=-\alpha^2dt^2 + \psi^4 [dr^2+r^2\!d\theta^2 + r^2\!\sin^2\!\theta (d\phi - \omega dt)^2],
\ee
with $\psi$ the {\it conformal factor}. Here we have assumed spherical-type coordinates $x^1=r$, $x^2=\theta$, $x^3=\varphi$, and $R^2 =  \psi^4 r^2\!\sin^2\theta$ in the present approximation. The Einstein equations then simplify to the system
\be
\Delta \psi = - [ 2\pi E+\textstyle{\frac{1}{8}} K_{ij}K^{ij} ] \,\psi^5,
\ee
\be
\Delta (\alpha\psi) = [ 2\pi (E+2S) + \textstyle{\frac{7}{8}} K_{ij}K^{ij} ] \,\alpha\psi^5,
\ee
\be
\Delta \omega = - 16\pi\alpha\psi^4 S^\phi - 2\psi^{10} K^{\phi j} \partial_j (\alpha\psi^{-6}),
\ee
where $\Delta$ is the standard Laplacian in spherical coordinates, $K_{ij}$ is the {\it extrinsic curvature}, computed via derivatives of $\omega$, whereas the sources $E$, $S$, and $S^\phi$ ($E$ is the total energy, $S^i$ is the total momentum density, $S$ is the trace of the total energy-momentum tensor) are determined from the fluid quantities and EM fields. However, in the rotating case, a more efficient and very robust method is the {\it eXtended} CFC approximation (XCFC) allowing for a hierarchical scheme and uniqueness of solution (one extra equation needed) \cite{Cordero-Carrion:2009}. The XCFC metric has been employed in our \texttt{ECHO} code for full GRMHD evolution \cite{Del-Zanna:2007} to extend it to dynamical spacetimes (\texttt{X-ECHO} \cite{Bucciantini:2011a}).

Here we briefly describe the \texttt{XNS} code, a version of \texttt{X-ECHO} specially designed to build accurate models for stationary and axisymmetric configurations of magnetized relativistic stars \cite{Pili:2014} (we could verify that models for rotating NSs are consistent with full GR within accuracy of $10^{-4}$). The numerical solution is found thanks to the iterative scheme:
\begin{enumerate}
\item provide an initial static and radially symmetric guess (Tolman-Oppenheimer-Volkoff solution),
\item solve Einstein equations for the XCFC metric $\, \to \alpha, \psi, \omega$,
\item solve Maxwell equations (or GS) for EM fields $\,\to\Psi, \Phi \, \to \mathbf{B}, \mathbf{E}$,
\item solve the Bernoulli integral for matter (any EoS) $\, \to \rho, p, \mathbf{v}$,
\end{enumerate}
to be stopped when the desired tolerance is reached. 

In the code the nonlinear PDEs for the metric and the electromagnetic potentials are solved via decomposition into scalar and vector spherical harmonics
\be
u(r,\theta) = \sum A_l(r) Y_l (\theta), \quad X^{\phi}(r,\theta) = \sum C_l(r) Y^\prime_l (\theta),
\ee
grid discretization, and direct inversion of tridiagonal matrices, where the scalar function $u$ is either $\alpha$, $\psi$, or $\Phi$, whereas $X^\phi$ refers to either $\omega$ or $\Psi$. As mentioned above, in the (uniformly) rotating case we assume an external vacuum solution for simplicity. The functions $\Phi$ and $\Psi$ are thus independently derived in the magnetosphere, while in the star interiors they must satisfy the ideal MHD condition $\Phi=-\Omega\Psi + C$, where $C$ is an integration constant proportional to the monopolar surface charge of the star. In addition, a special care must be ensured at the star surface in finding the potential $\Phi$, which must smoothly connect plasma and vacuum solutions \cite{Pili:2017}.

\subsection{Numerical models}

Let us now show a couple of examples of rotating equilibria which can be found by using the \texttt{XNS} code and the procedure described above, for sake of clarity one for purely toroidal fields and one for purely poloidal ones (though a combination of the two is also allowed). Here the reference star has a gravitational mass $M=1.55 M_\odot$, and a polytropic EoS $p=K\rho^2$, with $K=110$ in geometrized units.

\begin{figure}[t]
\centering
\includegraphics[height=65mm]{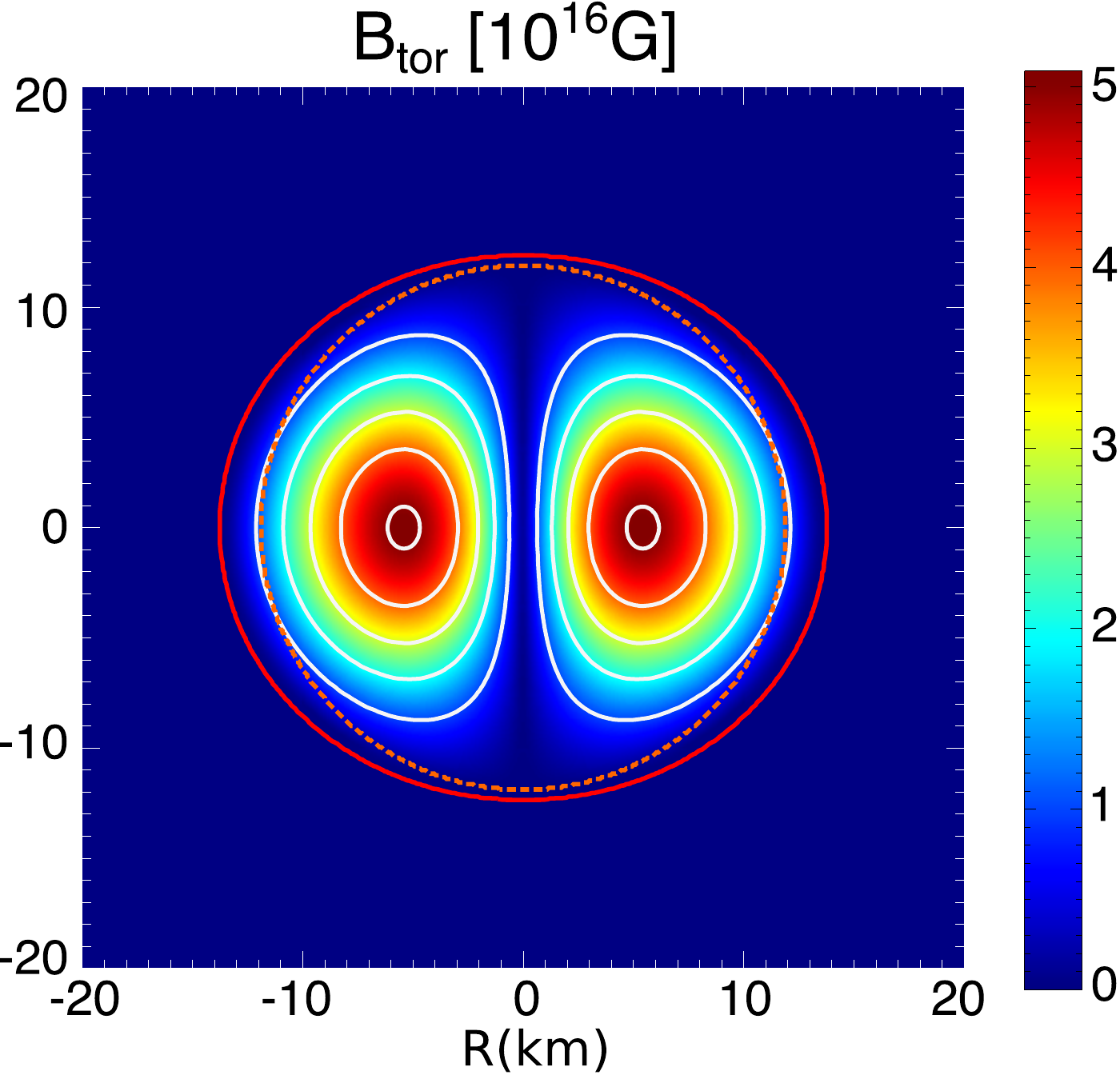} 
\caption{
A uniformly rotating magnetized neutron star with a purely toroidal field ($m=1$ case). The field strength $B_{\hat{\phi}}$ is displayed in colors. The external line represents the star surface, while the dashed line is that corresponding to the static and unmagnetized case (adapted from \cite{Pili:2017}).
}
\label{fig:NStor} 
\end{figure}

When the field has a vanishing poloidal component, we can safely assume $\Psi=0$ and the Grad-Shafranov equation does not come into play. Integrability of the equations requires instead
$$
\rho h \alpha^2 R^2\nabla\mathcal{M} + \mathcal{I}\nabla\mathcal{I} = 0,
$$
so that the two functions $\mathcal{I}$ and $\mathcal{M}$ are not independent, and it is convenient to choose
$$
\mathcal{I} = \alpha R B_{\hat{\phi}} = K_m (\rho h \alpha^2 R^2)^m,
$$
where $K_m$ is a constant and $m\geq 1$ \cite{Kiuchi:2008,Lander:2009,Frieben:2012}. In figure~\ref{fig:NStor} we show a configuration with rotation rate $\Omega = 3.05\times 10^3\,\mathrm{s}^{-1}$, maximum field $B_\mathrm{max}=5.1\times 10^{16}\,\mathrm{G}$, and $m=1$. Notice that in the toroidal case the magnetic field is contained inside the star and must vanish at the rotation axis. It is interesting to measure the so-called mean deformation of the star defined as
$$
\bar{e} = (I_{zz} - I_{xx})/I_{zz},
$$
since a change of inclination of the rotation axis in the first phases of a magnetar life is expected to induce a strong emission of gravitational waves provided $|\bar{e}|\sim 10^{-3}$ \cite{DallOsso:2009}. Notice that while the  centrifugal force induced by rotation tends to increase the equatorial radius of the star, yielding an oblate shape with $I_{zz}>I_{xx}=I_{yy}$ ($\bar{e}>0$), as in figure, a very strong toroidal field would be able to squeeze the star via the Lorentz force (hoop stresses) substantially providing a prolate shape ($\bar{e}<0$). For $B_\mathrm{max} \le 10^{17}\,\mathrm{G}$ we find
$$
\bar{e}  \simeq - 9\times 10^{-3} (B_\mathrm{max}/10^{17}\,\mathrm{G})^2 + 
3\times 10^{-3} (P/10\,\mathrm{ms})^{-2},
$$
thus the order of magnitude is the right one to produce substantial gravitational waves emission.

\begin{figure}[t]
\centering
\includegraphics[height=65mm]{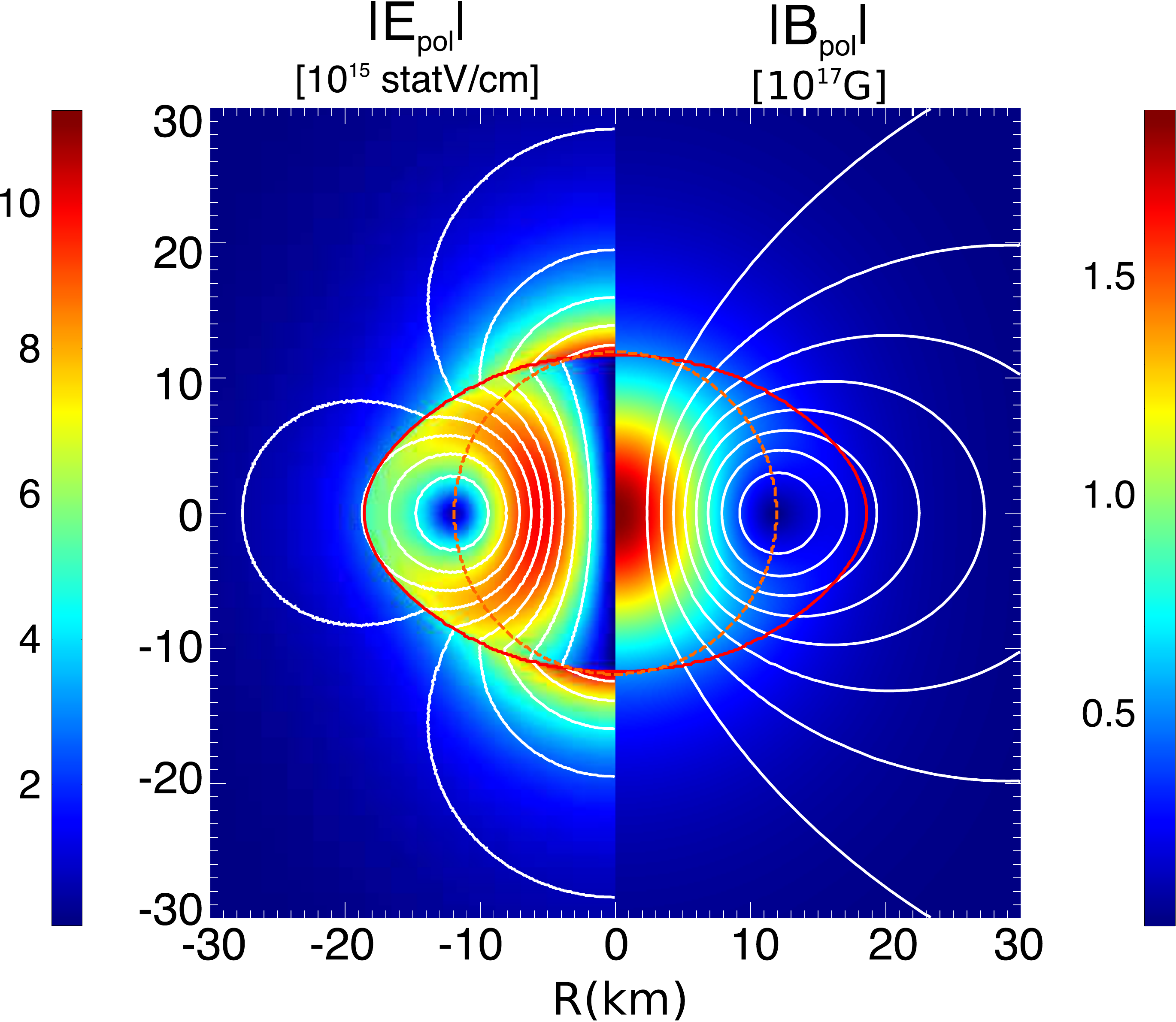} 
\caption{
A uniformly rotating magnetized neutron star with a purely poloidal field ($\xi=0$ case). In the left panel the electric field and the isocontours of the corresponding potential $\Phi$ are shown, while in the right panel we show the magnetic field and the isocontours of the corresponding potential $\Psi$. The external line represents the star surface, while the dashed line is that corresponding to the static and unmagnetized case (adapted from \cite{Pili:2017}).
}
\label{fig:NSpol} 
\end{figure}

Consider now the purely poloidal case, for which $\mathcal{I} = 0$. The shape of the magnetic field is selected by choosing the magnetization function $\mathcal{M}(\Psi)$. As in \cite{Ciolfi:2013,Bucciantini:2015} we allow for the nonlinear dependence
$$
\mathcal{M}(\Psi) = k_\mathrm{pol}\Psi \left(  1 + \frac{\xi}{|\nu+1|} \Psi^\nu  \right),
$$
though we restrict here to the linear case $\xi=0$ for simplicity, as often adopted in the literature \cite{Ciolfi:2009}. In figure~\ref{fig:NSpol} we show a configuration with rotation rate $\Omega = 4.06\times 10^3\,\mathrm{s}^{-1}$, and maximum field of $B_\mathrm{max}= 1.88 \times 10^{17}\,\mathrm{G}$. Here the magnetic fieldlines smoothly extend from the star interior to the outer magnetosphere, and in the present case we choose to have a vanishing global (monopolar) surface electric charge. Notice that inside the star the ideal MHD condition forces the electric field to be normal to the magnetic one, while in the magnetosphere the vacuum condition allows for polar regions where $\mathbf{E}\cdot\mathbf{B}\neq 0$, where particle extraction from the surface and acceleration is expected. The mean deformation for a relativistic star endowed with purely poloidal fields is always positive, thus yielding oblate configurations, and the effect increases with rotation approximately as
$$
\bar{e} \simeq 5\times 10^{-3} (B_\mathrm{max}/10^{17}\,\mathrm{G})^2 + 
3\times 10^{-3} (P/10\,\mathrm{ms})^{-2}.
$$

\section{Relativistic pulsar winds and their nebulae}

As discussed in the introduction, rotating and magnetized NSs accelerate an outflow of fields and particles from their magnetospheres. Beyond the light cylinder this wind is often modeled as a plasma outflow described by (special) relativistic MHD equations, and characterized by a certain degree of {\it magnetization}
$$
\sigma = \frac{B^2}{4\pi\rho c^2\gamma^2},
$$
namely the ratio of Poynting to fluid energy fluxes in the case of a mainly radial (cold) outflow with $v\to c$ with Lorentz factor $\gamma$ and a dominant toroidal field component $B\equiv B_{\hat{\varphi}}$. The wind magnetization is expected to be very high just outside the magnetosphere, where the FFE regime is valid, and to decrease in the acceleration region. When $\gamma$ is finally constant also $\sigma$ should not vary according to ideal MHD theory, though dissipation around the equatorial current sheet (the so-called {\it striped wind region}, that could be a large sector in the case of oblique rotators) strongly modify this picture \cite{Kirk:2009,Petri:2016a}.


In the case of young pulsars, the (still powerful) wind hits the surrounding slowly expanding ejecta, forming the PWN bubble through the TS, as anticipated in the introduction. Radially symmetric steady models \cite{Rees:1974,Kennel:1984} predict a strong field dissipation to occur before the TS, such as to reach $\sigma\sim 10^{-3}$ there (approximately $10^9$ light cylinder radii), though more realistic 2-D and 3-D simulations allow to relax this stringent constraint, so that $\sigma\sim 1$ at the TS or even larger in cases of wide striped wind regions \cite{Porth:2013}. 

\begin{figure}[t]
\centering
\includegraphics[height=65mm]{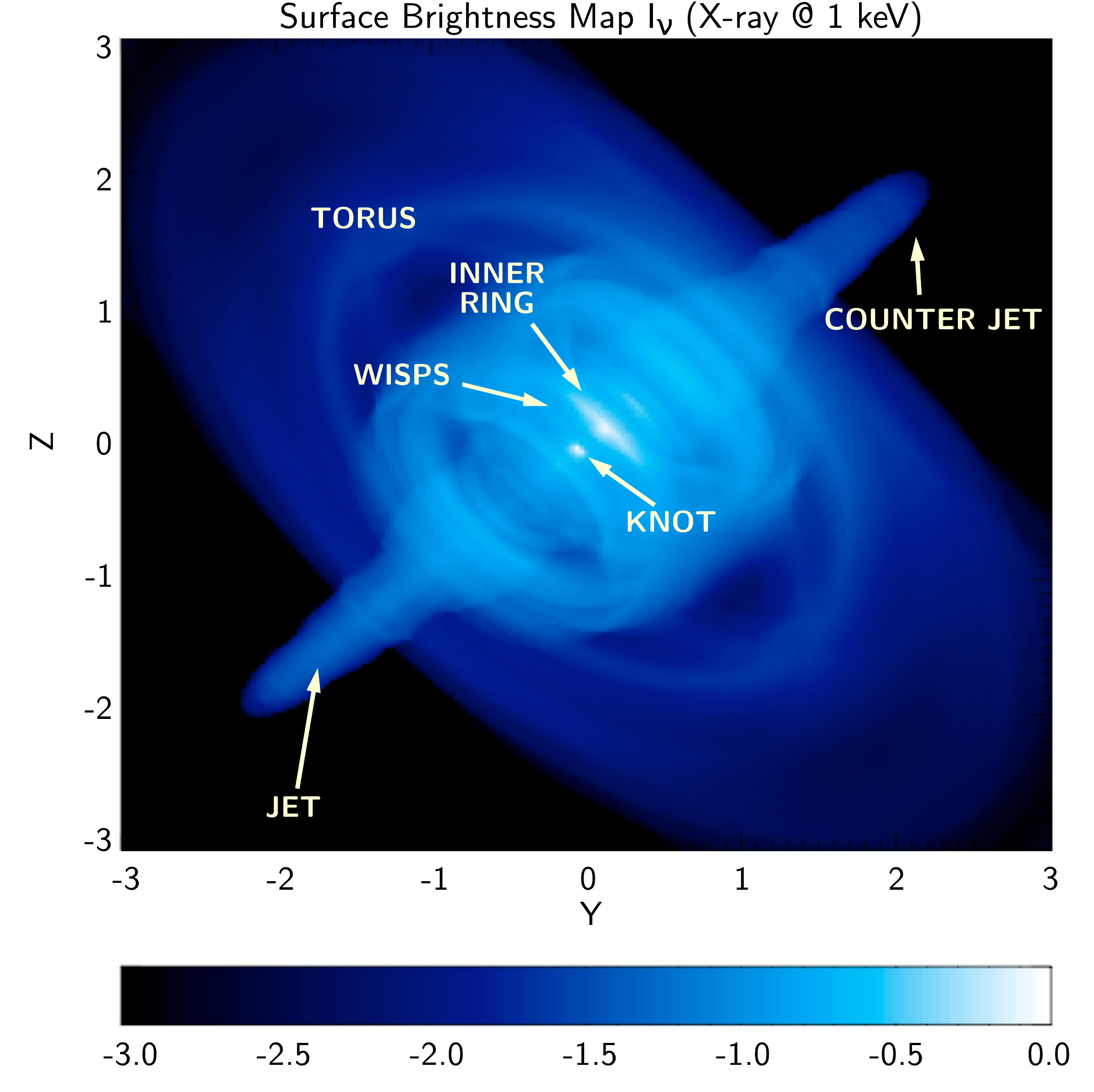} 
\caption{The simulated X-ray (at 1~keV) surface brightness map of the Crab nebula (adapted from \cite{Del-Zanna:2006}). A logarithmic scale is employed and distances are measured in ly.}
\label{fig:emission} 
\end{figure}

In spite of the success of steady 1-D models, also in terms of predictions on the non-thermal synchrotron and inverse Compton emission from the PWN \cite{Kennel:1984a,Atoyan:1996}, the high-resolution Chandra X-ray images of the Crab nebula \cite{Weisskopf:2000}, with an unexpected jet-torus inner structure of the PWN, imposed a substantial revision of MHD modeling. Moving to a more realistic axisymmetric view, the torus of enhanced emission could be explained by a pulsar wind with anisotropic energy flux, while the polar jets, which were previously thought to be originated at the pulsar location, could be accelerated thanks to hoop stresses by the amplified (toroidal) magnetic field downstream of the TS \cite{Bogovalov:2002,Lyubarsky:2002}. Thanks to the progresses in computational power and techniques, see \cite{Del-Zanna:2017} for a wide review on methods and numerical results, most of these issues could be settled. Time-dependent axisymmetric simulations in the (ideal) relativistic MHD regime were able to fully confirm this picture and to reproduce the inner structure of PWNe down to many fine details, such as the time-varying Doppler-boosted features known as the {\it knot} and the moving {\it wisps} \cite{Komissarov:2004a,Del-Zanna:2004,Del-Zanna:2006,Volpi:2008c,Camus:2009,Komissarov:2011,Olmi:2014,Olmi:2015}, as shown in figure~\ref{fig:emission}.

\begin{figure}[t]
\centering
 	\quad \includegraphics[height=75mm]{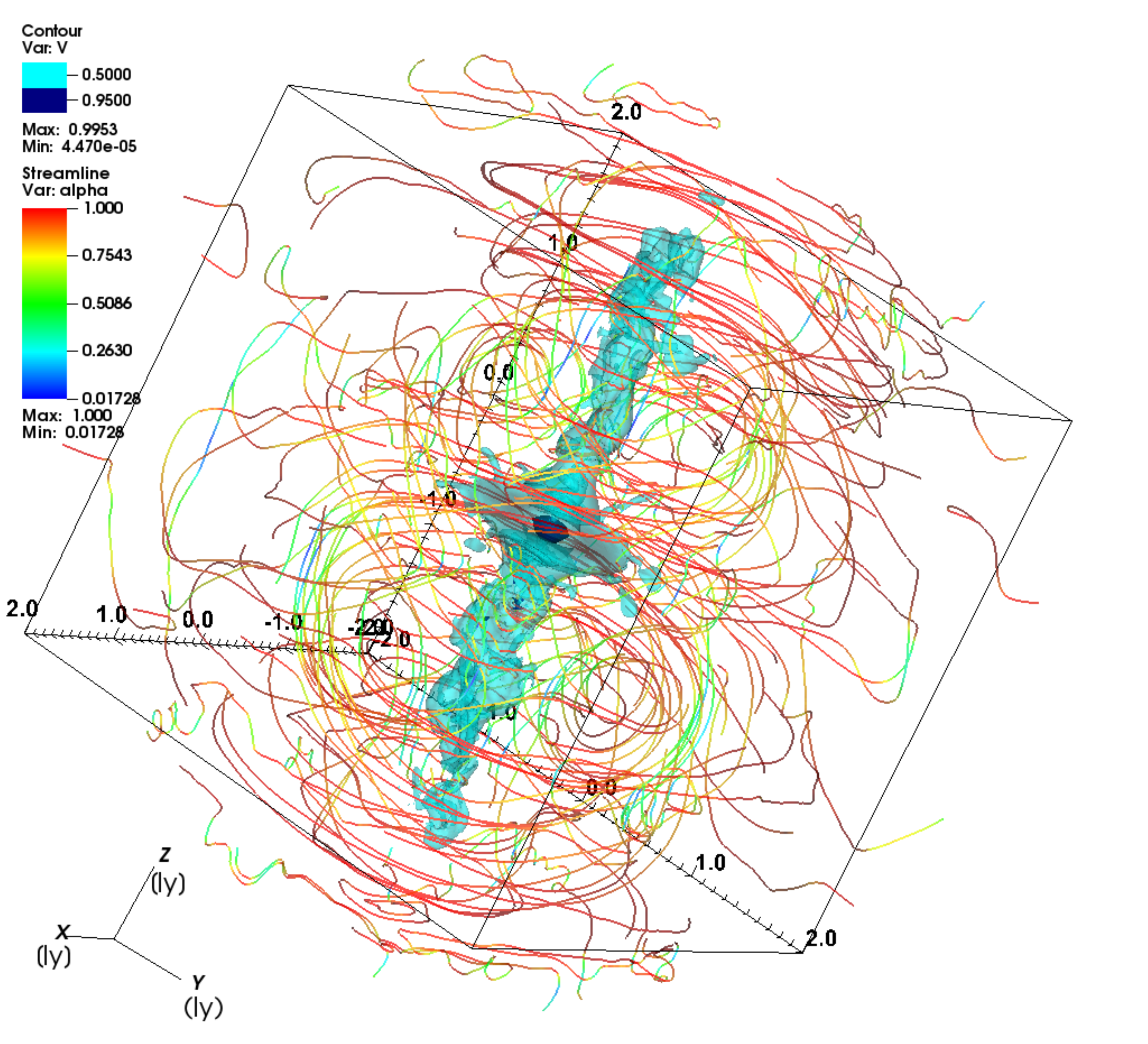}
\caption{Magnetic field lines and plasma velocity at the time $t=250$~y (from \cite{Olmi:2016}). The contrast between toroidal and poloidal components is indicated by the quantity $\alpha=B_\mathrm{tor}/B$ (red for a completely toroidal field, blue for a purely poloidal one). The flow speed is represented as a 3-D contour plot, with color levels corresponding to $0.95 c$ (basically at the TS) and $0.5 c$ (polar jets).} 
\label{fig:3dfield}
\end{figure}

Even if a purely toroidal field is evidently a good approximation for the inner PWN regions, given the success of axisymmetric models, in the outer nebula and along the polar axis, where jets are observed to kink \cite{Pavlov:2003}, this hypothesis must be abandoned leading to extra channels for magnetic field instability and dissipation \cite{Begelman:1998}. The first 3-D relativistic MHD simulations \cite{Porth:2013,Porth:2014,Olmi:2016} fully confirm this scenario, and magnetizations of order unity or more (averaged along the polar angle) at the TS can be reached, finally solving the long-standing $\sigma$ issue. While the inner structure of both plasma flow and magnetic field is very similar to the previous case, including the motion of wisps and the appearance of a luminous knot, a substantial fraction of magnetic energy now goes into the poloidal components in the outer regions and in jets, which are less strong compared to the 2-D case and are subject to the kink instability, as expected (see figure~\ref{fig:3dfield}). 

Unfortunately, 3-D simulations of PWNe are extremely demanding in terms of computing time, due to the large number of levels of mesh refinement needed to treat the evolution of ultra-relativistic pulsar wind region and of the TS. Because of this, it has been so far impossible to follow the entire evolution of the PWN (1000 years for the Crab nebula), and only recently the self-similar stage of PWN evolution has been reached \cite{Olmi:2016}. In this work, synthetic emission maps are obtained at radio, optical and X-ray frequencies. The synchrotron burn-off effect is observed through the decrease of the size of the emitting area with increasing observation frequency. We have also computed maps of the time variability at radio frequencies for both the wisps region (the inner nebula) and for the entire nebula, obtaining very encouraging results and enabling us to tune the spatial dependence of particle injection along the TS. Future plans are to address the issue of the integrated synchrotron and inverse Compton multi-wavelength spectrum, to better estimate also the number of injected particles. The final goal, probably close to be at reach, will be to derive all the unknown physical parameters of the pulsar wind, or even of the inner NS magnetosphere, by comparing the synthetic non-thermal emission from 3-D MHD models with the observational data.

Particular attention will be devoted to the dissipation of the nebular magnetic field, which has proven to be crucial in the solution of the $\sigma$ problem. Especially in full 3-D, where numerical resolution is necessarily low far from the TS for computational reasons, this has been certainly enhanced in these preliminary simulations by numerical effects due to coarseness of the grid. Full treatment of resistive relativistic MHD would be the best approach, even if  the necessary resolution of thin current sheets, where fast reconnection and magnetic dissipation is expected to take place \cite{Del-Zanna:2016a}, will be very hard to achieve. Moreover, another ingredient that is expected to affect the emission from the outer regions is certainly particle diffusion. This is largely enhanced by MHD turbulence and has proved to better reproduce the spectra of more evolved PWNe with both semi-analytical \cite{Tang:2012} and fully numerical (MHD and kinetic particles) models \cite{Porth:2016}. We leave these challenging tasks as future work.

\section{Acknowledgments}
The authors acknowledge support the PRIN-MIUR project prot. 2015L5EE2Y {\it Multi-scale simulations of high-energy astrophysical plasmas}, from the EU FP7-CIG grant issued to the NSMAG project, from the INFN-TEONGRAV initiative, and from the INFN-CIPE grant {\it High performance data network: convergenza di metodologie e integrazione di infrastrutture per HPC e HTC}.

\section*{References}

\bibliographystyle{iopart-num}


\providecommand{\newblock}{}

\end{document}